\documentstyle[twocolumn,prl,aps]{revtex}

\draft
\tighten

\begin{document}

\title{Density Matrix Perturbation Theory}

\author{Anders~M.~N.~Niklasson$^\dagger$  and Matt~Challacombe}

\address{Theoretical Division, Los Alamos National Laboratory,
Los Alamos, NM 87545, USA}

\date{\today}
\maketitle

\footnotetext{Preprint LA-UR 03-6452}

\begin{abstract}
{\small \bf An expansion method for perturbation of the zero temperature 
grand canonical density matrix is introduced.
The method achieves quadratically convergent recursions that yield
the response of the zero temperature density matrix upon variation 
of the Hamiltonian. The technique allows treatment of embedded 
quantum subsystems with a computational cost scaling linearly 
with the size of the perturbed region, ${\cal O}(N_{\rm pert.})$, 
and as ${\cal O}(1)$ with the total system size. It also allows 
direct computation of the density matrix response functions to 
any order with linear scaling effort. 
Energy expressions to 4th order based on only first and second 
order density matrix response are given.}
\\

\end{abstract}

In electronic structure 
theory, significant effort has been devoted to the development of
methods with the computational cost scaling linearly with system size 
\cite{Goedecker_RMP_99,Wu02}. The ability to perform accurate calculations
with reduced-complexity ${\cal O}(N)$ scaling is an important breakthrough 
that opens a variety of new possibilities in computational materials science, 
chemistry and biology. One of the most elegant and efficient approaches 
to linear scaling is density matrix purification 
\cite{McWeeny60,Clinton69,Palser98,Holas01,NiklassonWLT,NiklassonSP2,NiklassonSP4},
where constructing the density matrix by quadratically convergent spectral projections 
replaces the single-particle eigenvalue problem arising in tight-binding and 
self-consistent Hartree-Fock and Kohn-Sham theory. For large insulating systems 
this method is efficient because of the sparse ${\cal O}(N)$ real-space matrix 
representation of operators.  Instead of cubic scaling, the computational cost 
scales linearly with the system size.  Apart from ${\cal O}(N)$ purification 
techniques there are alternative approaches such as constrained functional 
minimization \cite{Li93,Kohn96}, and hybrid schemes
\cite{Challa99,Bowler99,Daniels99,Helgaker2000,Head_Gordon2003}. 

In this letter, we introduce a grand canonical density matrix perturbation theory based
on recently developed spectral projection methods for purification of the density matrix
\cite{NiklassonSP2,NiklassonSP4}. The method provides direct solution of the zero temperature 
density matrix response upon variation of the Hamiltonian through quadratically convergent 
recursions. The method makes it possible to study embedded quantum subsystems 
and density matrix response functions within linear scaling effort. The density matrix 
perturbation technique avoids using wavefunction formalism.
In spirit, it is therefore similar to the density matrix perturbation method
proposed by McWeeny \cite{McWeeny_PRT} and offers a flexibility comparable to
Green's function methods \cite{Haydock80,Inglesfield81}.
The present work is likewise related to the recent work of Bowler and Gillan \cite{Bowler02},
who developed a functionally constrained density matrix minimization scheme for embedding.
However, our approach to computation of the density matrix response is quite different from 
existing methods of solutions for the coupled-perturbed self-consistent-field equations.  
In contrast to conventional methods that pose solution implicitly through coupled equations 
\cite{Frisch,Dupuis,Ochsenfeld,Larsen}, the new method provides explicit construction of the 
derivative density matrix through recursion.  

The main problem in constructing a
density matrix perturbation theory is the non-analytic
relation between the zero temperature density matrix and 
the Hamiltonian, given by the discontinuous step function \cite{Notation},
\begin{equation} \label{DM}
P = \theta[\mu I -  H].
\end{equation}
This discontinuity makes expansion of $P$ about $H$ difficult.
At finite temperatures the discontinuity disappears and we may use perturbation 
expansions of the analytic Fermi-Dirac distribution \cite{Feynman}. 
However, even at finite temperatures a perturbation expansion based
on the Fermi-Dirac distribution may have slow convergence.

In linear scaling purification schemes 
\cite{McWeeny60,Clinton69,Palser98,Holas01,NiklassonWLT,NiklassonSP2,NiklassonSP4}, 
the density matrix is constructed by recursion;
\begin{equation}\label{DM_EXP} \begin{array}{ll}
X_0     & = L(H), \\
X_{n+1} & = F_n(X_n), ~~ n = 0,1,2, \ldots\\
P    & = \lim_{n \rightarrow \infty} X_n. \end{array}
\end{equation}
Here, $L(H)$ is a linear normalization function mapping all eigenvalues of $H$
in reverse order to the interval of convergence $[0,1]$ and $F_{n}(X_n)$ 
is a set of functions projecting the eigenvalues of $X_n$
toward  1 (for occupied states) or 0 (for unoccupied states). In one of 
the simplest and most efficient techniques, which requires only 
knowledge of the number of occupied states $N_e$ and no {\em a priori} 
knowledge of $\mu$ \cite{NiklassonSP2}, we have 
\begin{equation} \label{SP2}
F_{n}(X_n) = 
\left\{\begin{array}{ll}
X_n^2, &  Tr(X_n) \geq N_e \\
2X_n - X_n^2, & Tr(X_n) < N_e.
\end{array} \right.
\end{equation}
Purification expansion schemes are quadratically convergent,
numerically stable, and can even solve problems with degenerate
eigenstates and fractional occupancy \cite{NiklassonSP4}. Thanks to an 
exponential decay of the density matrix elements as a function of $|{\bf r-r'}|$ for 
insulating materials, the operators have a sparse matrix 
representation and the number of non-zero matrix elements above a numerical threshold
scales linearly with the system size.  In these cases the matrix-matrix multiplications, 
which are the most time consuming steps, have an $N$-scaling cost.

Equivalent to the purification
schemes are the sign-matrix expansions \cite{Kenney91,Beylkin99,Karoly}.
The general scheme is the same as in Eq.\ (\ref{DM_EXP}), but
the expansion is performed around a step from $-1$ to $1$ at $x=0$.

Our grand canonical density matrix perturbation theory is based on 
the purification in Eq.\ (\ref{DM_EXP}).
A perturbed Hamiltonian $H = H^{(0)} + H^{(1)}$ gives
the expansion 
\begin{equation}\label{dX}
{X_n} = X_n^{(0)} + \Delta_n, ~ n = 0,1,2,\ldots~,
\end{equation}
where $X_n^{(0)}$ is the unperturbed expansion 
and $\Delta_n$ are the differences due to the perturbation $H^{(1)}$. 
It is then easy to show that 
\begin{equation}\label{DYS_F}\begin{array}{ll}
\Delta_{n+1} &= F_n(X_n^{(0)} + \Delta_n) - F_n(X_n^{(0)})\\
{P} &= P^{(0)} + \lim_{n \rightarrow \infty} \Delta_n. \end{array}
\end{equation}
This is the key result of the present
article and defines our grand canonical density matrix perturbation theory.
Combining Eq.\ (\ref{DYS_F}) with the expansion in Eq.\ (\ref{SP2}) gives 
the recursive expansion \cite{Notation}
\begin{equation} \label{DYS_SP2}
\Delta_{n+1} = 
\left\{\begin{array}{l}
\{ X_n^{(0)},\Delta_n\} + \Delta_n^2 ~~~{\rm if} ~~  Tr(X_n^{(0)}) \geq N_e \\
2\Delta_n - \{ X_n^{(0)},\Delta_n\} - \Delta_n^2 ~~ {\rm otherwise}.
\end{array} \right.
\end{equation}
Other expansions based on, for example, McWeeny, trace conserving or 
trace resetting purification \cite{McWeeny60,Palser98,NiklassonSP4}
can also be included in this quite general approach. However, 
Eq.\ (\ref{DYS_SP2}) is particularly efficient since it only 
requires two matrix multiplications per iteration.
Because the perturbation expansions inherit properties from their 
generator sequence, they are likewise quadratically convergent with 
iteration, numerically stable, and exact to within accuracy 
of the drop tolerance \cite{NiklassonSP4}.

If the perturbed ${X}_0^{(0)}$ has eigenvalues outside the interval 
of convergence $[0,1]$ the expansion could fail. To avoid this problem 
the normalization function $L(H)$ in Eq.\ (\ref{DM_EXP}) can be chosen to contract the
eigenvalues of $X_0$ to $[\delta,1-\delta]$, where $\delta>0$ is sufficiently large.

A major advantage with the expansion in Eq.\ (\ref{DYS_SP2}) is that for band-gap 
materials that are locally perturbed, the $\Delta_n$ are likewise localized as a result 
of nearsightedness \cite{Kohn59,Kohn96}. The matrix products in Eq.\ (\ref{DYS_SP2}) 
can therefore be calculated using only the local regions of $X_n$ that respond to the perturbation.
Given that perturbation does not change the overall decay of the
density matrix, the computational cost of the expansion scales linearly with the
size of the perturbed region ${\cal O}(N_{\rm pert.})$ and as ${\cal O}(1)$ with
the total system size.

Density matrix purification does not necessarily require
prior knowledge of the chemical potential, but once the
initial expansion of the unperturbed system is carried out, the
chemical potential is set. The perturbation expansions of 
Eq.\ (\ref{DYS_F}) are therefore grand canonical \cite{CPRT}.
With this in mind, Eq.~(\ref{DYS_SP2}) may be readily applied 
to embedding schemes that do involve long range charge flow.

The computation of many spectroscopic properties such as the Raman spectra, 
chemical shielding and polarization requires the calculation of density 
matrix derivatives with respect to perturbation.
Grand canonical density matrix perturbation 
theory can be used to compute these response functions.
Assume a perturbation of the Hamiltonian $H^{(0)}$,
\begin{equation}
{H} = H^{(0)} + \lambda H^{(1)},
\end{equation}
in the limit $\lambda \rightarrow 0$.  
The corresponding density matrix is
\begin{equation}
{P} = P^{(0)} + \lambda P^{(1)} + \lambda^2 P^{(2)} + \ldots~,
\end{equation}
where the response functions $P^{(\mu)}$ (density matrix derivatives) correspond 
to order $\mu$ in $\lambda$.  Expanding the perturbation as in 
Eq. (\ref{DYS_SP2}), individual response terms may be collected
order by order at each iteration;
\begin{equation}
\Delta_n = \lambda \Delta^{(1)}_n + \lambda^2 \Delta^{(2)}_n + \ldots~.
\end{equation}
Keeping terms through order $m$ in $\lambda$ at each iteration, 
with $\Delta^{(0)}_n = X_n$, the following recursive sequence is obtained 
for $\mu = m,m-1,\ldots,1$ :
\begin{eqnarray}\label{second1}
\Delta^{(\mu)}_{n+1} = \left\{ \begin{array}{l}
\sum_{i=0}^{\mu} \Delta^{(i)}_n \Delta^{(\mu-i)}_n\quad {\rm if} ~~ {Tr}(X_n) \geq N_e\\ 
2\Delta^{(\mu)}_n - \sum_{i=0}^{\mu} \Delta^{(i)}_n \Delta^{(\mu-i)}_n \quad {\rm otherwise.}
\end{array} \right.
\end{eqnarray}
These equations provide an explicit, quadratically convergent solution of the response functions, 
where 
\begin{equation}
P^{(\mu)} = \lim_{n \rightarrow \infty} \Delta^{(\mu)}_n.  
\end{equation}
With the same technique it is possible to treat perturbations where 
${H} = H^{(0)} + \lambda_a H^{(1)}_a+\lambda_b H^{(1)}_b+ 
\lambda_a \lambda_b H^{(2)}_{a,b} + \ldots$
to produce a mixed density matrix expansion
${P} = P^{(0)} + \lambda_a P_a^{(1)} + \lambda_b P_b^{(1)} + \lambda_a \lambda_b P_{a,b}^{(2)} + \ldots~$.

Equation~(\ref{second1}) provide direct explicit construction of the response 
equations based on well developed linear scaling technologies \cite{NiklassonSP2,NiklassonSP4}.  
This is quite different from conventional approaches\cite{Frisch,Dupuis,Ochsenfeld,Larsen}  
that pose solution implicitly through coupled matrix 
equations, achieving at best linear scaling with iterative solvers.

Higher order expansions of an observable can be calculated efficiently from low
order density matrix terms. Similar to Wigner's 2n+1 rule for
wavefunctions \cite{Helgaker} we have the energy response 
${E} = E^{(0)} + \lambda E^{(1)} + \lambda^2 E^{(2)} + \lambda^3 E^{(3)} 
+ \lambda^4 E^{(4)}$, where
\begin{eqnarray}
E^{(1)} &=& Tr(P^{(0)} H^{(1)}), ~ E^{(2)} = 0.5 Tr(P^{(1)} H^{(1)}) \nonumber \\
E^{(3)} &=& Tr([P^{(1)},P^{(0)}] P^{(1)} H^{(1)}), \\
E^{(4)} &=& 0.5 Tr([(2I-P^{(0)}) P^{(2)} P^{(0)} P^{(1)} \nonumber \\
&~& - P^{(0)} P^{(1)} P^{(2)} (I+P^{(0)})] H^{(1)}). \nonumber
\end{eqnarray}
The corresponding n+1 rule for $\mu > 0$ is 
\begin{equation}
E^{(\mu)} = \mu^{-1} Tr(P^{(\mu-1)}H^{(1)}).
\end{equation}

To demonstrate the grand canonical density matrix
perturbation theory, we present two examples based on single-site 
perturbations of a model Hamiltonian and a beta-carotene molecule.

The model Hamiltonian has random
diagonal elements exhibiting exponential decay of the overlap elements 
as a function of site separation on a randomly distorted 
lattice. This model represents a Hamiltonian of an insulator
that might occur, for example, with a Gaussian basis set in 
density functional theory or in various tight-binding schemes. 
A local perturbation is imposed on the model Hamiltonian
by moving the position of one of the lattice sites.
Using the perturbation expansion of Eq.\ (\ref{DYS_SP2}),
a series of perturbations $\Delta_n$ is generated. In each 
step a numerical threshold $\tau = 10^{-6}$ is applied as described in
\cite{NiklassonSP4}.  The lower inset in Fig.  \ref{DX} shows 
the number of elements above the threshold in $\Delta_n$ 
as a function of iteration. The local perturbation
is efficiently represented with only $\sim 50$ elements out of $10^4$.
Figure \ref{DX} also illustrates the quadratic convergence
of the error.  At convergence after 
$M$ iterations the new perturbed density matrix is given by
${P} = P^{(0)} + \Delta_M$. The error 
$||{P} - P_{\rm exact}||_2 = 6.4 \times 10^{-5}$
and the error $|{E} - E_{\rm exact}| = 1.3 \times 10^{-6}$
\cite{first_order}.  The error of the perturbed density matrix 
${P}$ is stable at convergence and close to the numerical error 
for the unperturbed density matrix due to thresholding
$||P^{(0)} - P^{(0)}_{\rm ~exact}||_2 = 1.0 \times 10^{-4}$, and
$|{E}^{(0)} - E^{(0)}_{\rm exact}| = 2.4 \times 10^{-6}$.

The electronic structure of the second example, the beta-carotene molecule, 
was calculated with the MondoSCF suite of linear scaling algorithms \cite{Mondo_SCF} 
at the RHF/STO-2G level of theory.  Figure \ref{BetaC} shows the matrix sparsity 
factor of the density matrix for beta-carotene. The difference between two fully 
self-consistent Fockians was chosen as a perturbation, 
(one with and one without a small displacement of a single carbon atom). 
In this way, more long-ranged effects due 
to self-consistency are included. Even if beta-carotene
is too small to have a very sparse representation of the density matrix, 
the perturbation sequence $\Delta_n$ is found to be highly sparse.
The error with threshold $\tau = 10^{-5}$ in the single-particle Hartree-Fock energy 
$|{E} - E_{\rm exact}| = 2.8 \times 10^{-5}$ a.u., which is of the
same order of error as for the unperturbed molecule. Standard first
order perturbation theory yields an error two orders of magnitude 
larger \cite{first}.

The present formulation has been developed in an orthogonal representation.  With 
a $N$-scaling congruence transformation \cite{Challa99},  it is straightforward to 
employ this representation when using a non-orthogonal basis. When using a 
non-orthogonal basis set, change in the inverse overlap matrix  $S^{-1}$  due to
a local perturbation $dS$ is given by the recursive Dyson equation, 
\begin{equation}
\delta_{n+1} = {S_{0}}^{-1} dS ({S_{0}}^{-1} + \delta_n),
\end{equation}
where $S = S_{0} - dS$, $\delta_{0} = 0$, and
$S^{-1} = {S_{0}}^{-1}+\lim_{n \rightarrow \infty} \delta_n$.
The equation contains only terms with local sparse updates 
and the computational cost scales linearly, ${\cal O}(N_{\rm pert.})$,
with the size of the perturbed region. Similar perturbation schemes 
for the sparse inverse Cholesky or square root factorizations can also 
be applied \cite{unpubl}.

Density matrix perturbation theory can be applied in many contexts. 
For example, a straightforward calculation of the energy difference 
due to a small perturbation of a very large system may not be possible because
of the numerical problem in resolving a tiny energy difference between
two large energies. With density matrix perturbation theory, 
we work directly with the density matrix difference $\Delta_n$ and the
problem can be avoided, for example, the single particle energy change
$\Delta E = \lim_{n \rightarrow \infty} Tr(H\Delta_n)$. In analogy to incremental Fock builds
in self-consistent field calculations \cite{Schwengler97}, the technique 
can be used in incremental density matrix builds.
Connecting and disconnecting individual weakly interacting \cite{Long_range}
quantum subsystems can be performed by treating off-diagonal elements of the
Hamiltonian as a perturbation. This should be highly useful in nanoscience 
for connecting quantum dots, surfaces, clusters and nanowires, where the different 
parts can be calculated separately, provided a connection through a common
chemical potential is given, for example via a surface substrate.
In quantum molecular dynamics, such as quantum mechanical-molecular mechanical QM/MM 
schemes, or Monte-Carlo simulations, where only a local part of the system is perturbed 
and updated, the new approach is of interest. 
Several techniques used within the Green's function context also should 
apply for the density matrix. The proposed perturbation approach may 
be used for response functions, impurities, 
effective medium and local scattering techniques \cite{Haydock80,Inglesfield81,Turek,Igor}.  
The theory of grand canonical density matrix perturbation is thus a rich field 
with applications in many areas of materials science, chemistry and biology.

In summary, we have introduced a grand canonical perturbation theory 
for the zero temperature density matrix, extending quadratically convergent 
purification techniques to expansions of the density matrix upon variation
of the Hamiltonian.  The perturbation method allows the local adjustment of 
embedded quantum subsystems with a computational cost that scales as ${\cal O}(1)$
for the total system size and as ${\cal O}(N_{\rm pert.})$ for the 
region that respond to the perturbation, as demonstrated in Figs.~\ref{DX} 
and \ref{BetaC}. A new quadratically convergent $N$-scaling recursive approach 
to computing density matrix response functions has been outlined, and
energy expressions to 4th order in terms of only first and second order density matrix response 
were given.  The density matrix perturbation technique is surprisingly simple and 
offers an efficient alternative to several Green's function methods and conventional 
schemes for solution of the coupled perturbed self-consistent-field equations.

Discussions with  E.\ Chisolm, S.\ Corish, S.\ Tretiak, C.\ J.\ Tymczak, V.\ Weber, 
and J.\ Wills are gratefully acknowledged.

\begin{figure}
\caption{\small  The 
${\rm Error} = \log_{10}(||X^{(0)}_n + \Delta_n - P_{\rm exact}||_2)$
as a function of iterations $n$ ($N = 100$, $N_e = 50$). The lower 
inset shows the number of non-zero matrix elements of $\Delta_n$ 
above threshold $\tau = 10^{-6}$.  The upper inset shows the density matrix
perturbation.
\label{DX}}
\end{figure}

\begin{figure}
\caption{\small  The matrix sparsity factors (number of non-zero elements 
over the total number of elements) for beta-carotene (RHF/STO-2G). 
\label{BetaC}}
\end{figure}


\begin{references}
\bibitem{CPRT} A canonical perturbation theory can be constructed
by allowing the spectral projections $F_n(X_n+\Delta_n)$ to differ
from $F_n(X_n)$ in Eq.\ (\ref{DYS_F}), however, in this case the
locality of the expansion for a local perturbation is lost.
\bibitem{first} $|Tr((F^{(0)}+\Delta V)P^{(0)})-E_{\rm exact}| = 4.5 \times 10^{-3}$ a.u.,
where $F^{(0)}$ is the unperturbed Fockian.
\bibitem{Long_range} In the case of strong interaction and self-consistent calculations, 
long ranged coupling may occur from charge fluctuations and changed dipole field. 
In this case the perturbation is no longer local.
\end{references}
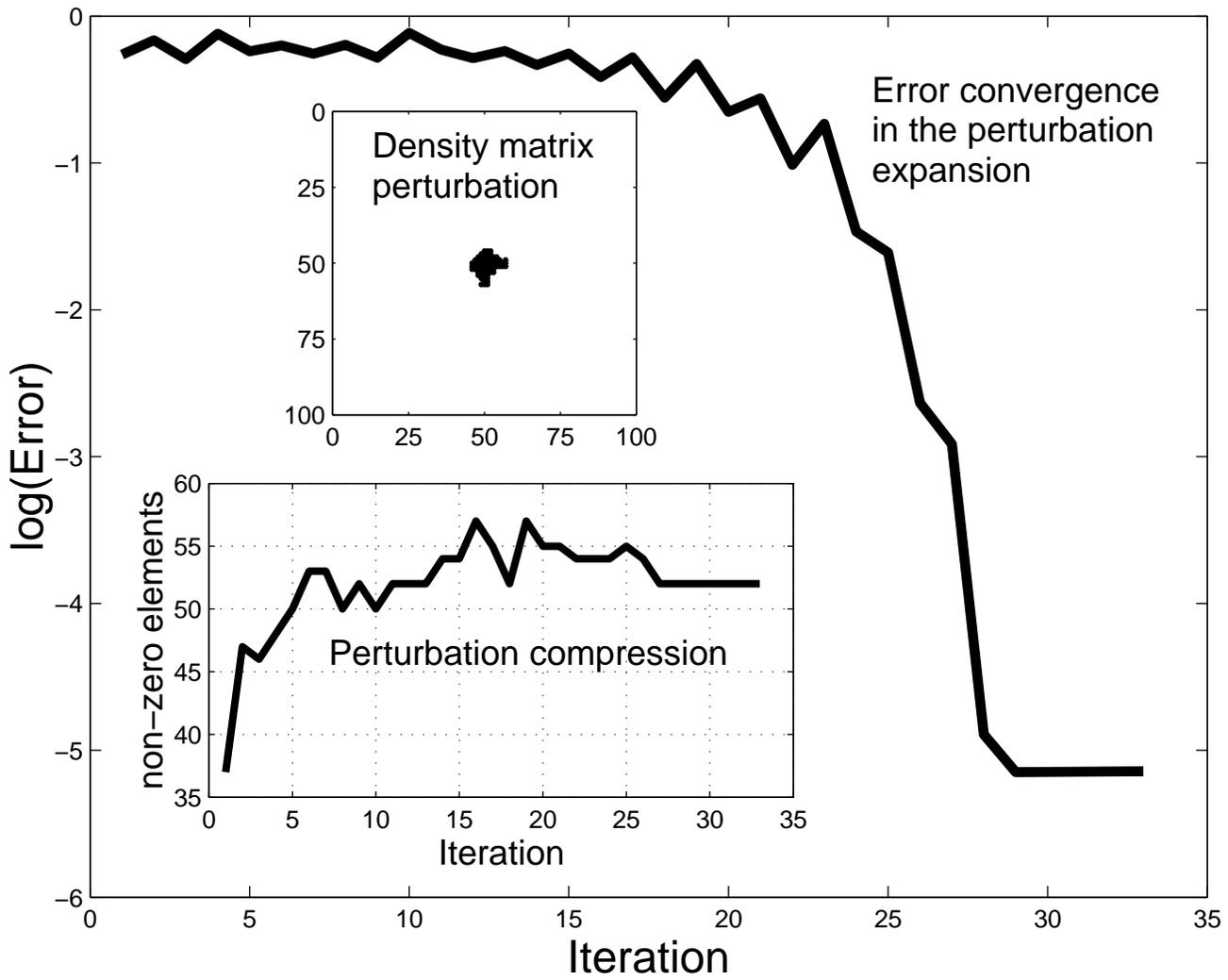
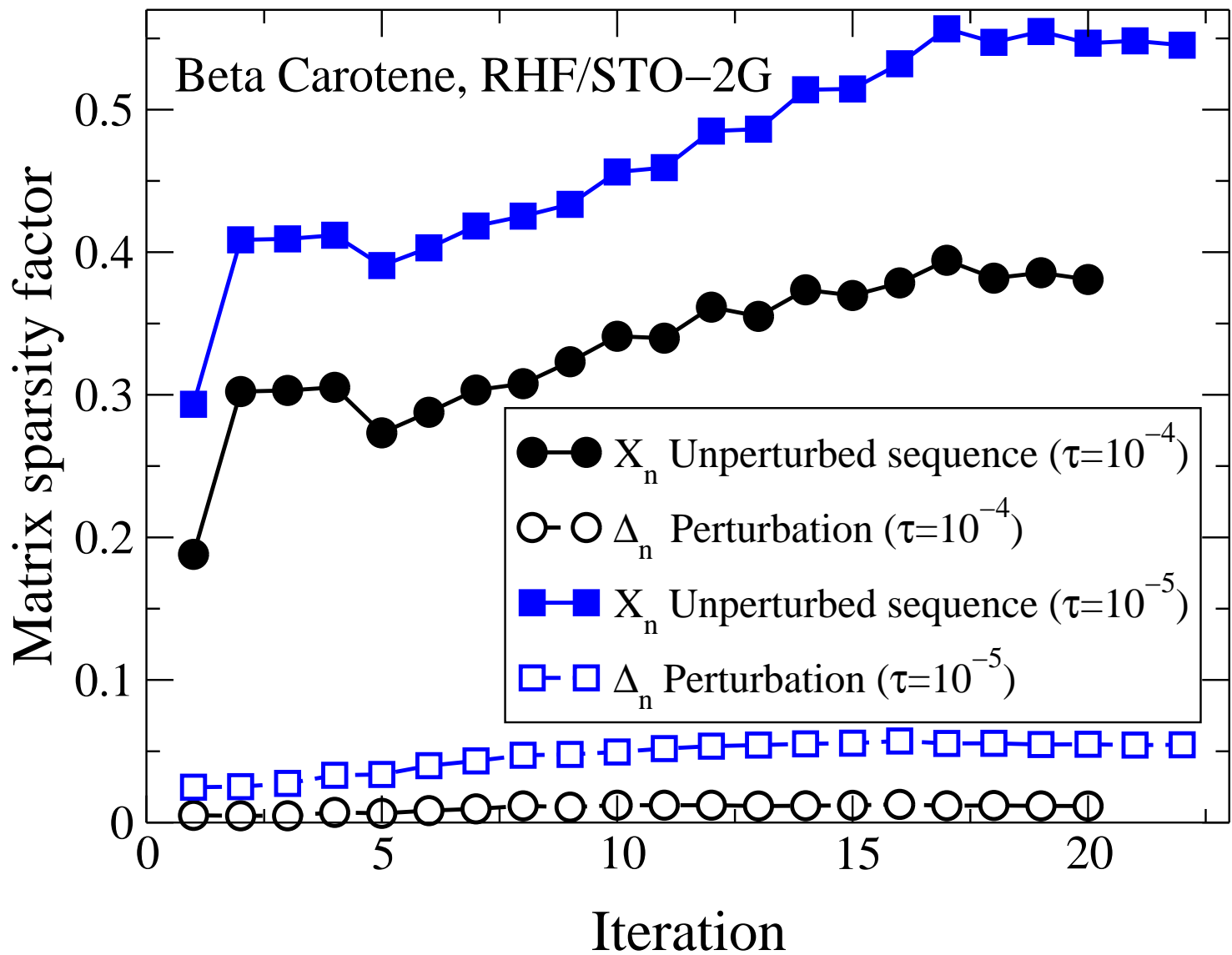
\end{document}